\pdfoutput=1

\documentclass[11pt,british]{article}
\usepackage[T1]{fontenc}
\usepackage[latin9]{inputenc}
\usepackage[a4paper]{geometry}
\usepackage{babel}
\usepackage{amstext}
\usepackage{amssymb}
\usepackage{graphicx}
\usepackage{esint}
\usepackage[unicode=true,
 bookmarks=true,bookmarksnumbered=false,bookmarksopen=false,
 breaklinks=false,pdfborder={0 0 1},backref=false,colorlinks=false]
 {hyperref}
\hypersetup{pdftitle={The cosmological constant filter without big bang singularity},
 pdfauthor={Florian Bauer},
 pdfsubject={No initial singularity in the Palatini filter model for large cosmological constants},
 pdfkeywords={Big bang singularity, cosmic bounce, cosmological constant, vacuum energy, modified gravity, Palatini formalism}}
\begin{document}

\title{The cosmological constant filter without big bang singularity}

\author{\textbf{Florian Bauer}\vspace*{0.5cm}
 \\
{\small High Energy Physics Group, Dept.\ ECM, and Institut de Ci\`encies
del Cosmos}\\
{\small{} Universitat de Barcelona, Av.\ Diagonal 647, E-08028 Barcelona,
Catalonia, Spain}\\
{\small Email: }\texttt{\small fbauerphysik@eml.cc}}

\date{{ }}
\maketitle
\begin{abstract}
In the recently proposed cosmological constant~(CC) filter mechanism
based on modified gravity in the Palatini formalism, gravity in the
radiation, matter and late-time de~Sitter eras is insensitive to
energy sources with the equation of state $-1$. This implies that
finite vacuum energy shifts from phase transitions are filtered out,
too. In this work we investigate the CC filter model at very early
times. We find that the initial big bang singularity is replaced by
a cosmic bounce, where the matter energy density and the curvature
are finite. In a certain case this finiteness can be observed already
on the algebraic level.
\end{abstract}

\section{Introduction}

The core of the old CC problem \cite{Weinberg:1988cp,Padmanabhan:2002ji,Bertolami:2009nr}
is the huge hierarchy between the observed value~$\Lambda_{0}\sim10^{-47}\,\text{GeV}^{4}$
of the CC in the $\Lambda$CDM model and the value~$\Lambda$ inferred
from theory. While the latter is difficult to calculate, it is usually
expected to be dominated by the ultraviolet sector of the theory implying
an enormous magnitude $|\Lambda|\gg\Lambda_{0}$. For example, the
most reliable contribution to~$\Lambda$ comes from the electro-weak
phase transition giving rise to the relation $|\Lambda|\sim(10^{2}\:\text{GeV})^{4}\sim10^{55}\Lambda_{0}$
\cite{Bauer:2010wj}. A common way to deal with this problem is adding
a CC counterterm, which subtracts off the huge vacuum energy density
and permits a reasonable low-energy universe. The price to pay is
the enormous finetuning of the counterterm, i.e.\ the precise choice
of tens of digits, which is an undesired property for any theory.
It is important to note that \emph{replacing} the CC by a dynamical
dark energy source generally does not avoid the old CC problem~\cite{Bauer:2010dg}.
However, it may help modelling the late-time accelerated expansion
and other cosmological epochs~\cite{Carroll:2000fy,Peebles:2002gy,Nojiri:2006ri,Copeland:2006wr,Elizalde:2009gx,Durrer:2011gq,Li:2011sd,SaezGomez:2011ny}.

Instead of removing the large CC by a static counterterm, we consider
a more dynamical approach to neutralise the effects of~$\Lambda$.
It is easy to see that in standard general relativity the large CC
would dominate the evolution of the universe at early times. Hence,
one has to introduce nontrivial changes in gravity \cite{Bauer:2009ea,Bauer:2010wj,Sola:2011qr,Bauer:2011em}
or in the matter sector \cite{Stefancic:2008zz,Bauer:2009ke,Bauer:2009jk}
in order to permit a low curvature cosmos. Further ways to address
the old CC problem can be found in Refs.~\cite{Bonanno:2001hi,Nobbenhuis:2004wn,Barr:2006mp,Diakonos:2007au,Klinkhamer:2007pe,Batra:2008cc,Charmousis:2011bf},
where, for instance, quantum gravity effects or self-adjustment mechanisms
are important. Finally, neutralising the effect of vacuum energy was
investigated in Refs.~\cite{Dvali:2007kt,Erdem:2007vh,Patil:2008sp,Demir:2009be,Demir:2011gi,Burgess:2011va}
from different perspectives. 

In this work we continue exploring the CC filter model proposed in
Ref.~\cite{Bauer:2010bu}. This approach to solve the old CC problem
is based on~$f(R,Q)$ modified gravity in the Palatini formalism~\cite{Vollick:2003ic,Sotiriou:2005xe,Barausse:2007ys,Sotiriou:2008rp,Olmo:2011uz},
where the gravity action is formulated in terms of the Ricci scalar~$R$
and the squared Ricci tensor~$Q$ \cite{Allemandi:2004wn,Li:2007xw,Li:2008bma,Li:2008fa}.
Palatini theories have a more involved algebraic structure than the
corresponding theories in the metric formalism~\cite{Exirifard:2007da,Bauer:2008zj,Borunda:2008kf,Capozziello:2009nq,DeFelice:2010aj,Amendola:2010bk,Vitagliano:2010sr,Harko:2010hw,Bauer:2010jg},
which permits the construction of a filter for vacuum energy. Moreover,
in our setup we find only second order field equations just as in
general relativity, thus extra degrees of freedom are avoided.

So far, only the region where the matter energy density $\rho$ is
smaller in magnitude than the CC~$\Lambda$ has been studied, which
includes the radiation, matter and late-time de~Sitter epochs. We
found already on the algebraic level that energy sources with the
equation of state~(EOS) $-1$ do not contribute to the curvature
at leading order, which is just a manifestation of the filter effect.
In the following we will extend the analysis by discussing the very
early cosmic epoch where $\rho$ may be of the order of $\Lambda$
or larger. As the main result we find that the filter model avoids
the occurrence of an initial curvature singularity, which is instead
replaced by a cosmic bounce with finite matter energy density and
finite curvature. For a certain case, this finiteness can be seen
also on the algebraic level. Finally, it is worth mentioning that
the absence of an initial singularity has been observed also for other
Palatini models in Refs.~\cite{Barragan:2010qb,Olmo:2011sw}, and
a connection to loop quantum cosmology was suggested~\cite{Olmo:2008nf}.

The paper is organised as follows: in Sec.~\ref{sec:fRQ-Palatini}
we briefly review how to solve the gravity field equations in the
Palatini formalism, and the procedure will be applied to the CC filter
model in Sec.~\ref{sec:CC-Filter}. We solve the main equations and
discuss the cosmological solutions in Sec.~\ref{sec:Solutions} before
we conclude in Sec.~\ref{sec:Conclusions}.

In this work the speed of light~$c$ and the Planck constant~$\hbar$
are set to unity, the signature of the metric is $(-1,+1,+1,+1)$.
The tensor indices~$a,\dots,f,m,n$ run from~$0$ to~$3$, whereas~$i,j=1,2,3$
are related to spatial coordinates.

\section{$f(R,Q)$ modified gravity in the Palatini formalism}

\label{sec:fRQ-Palatini}The CC filter mechanism is based on $f(R,Q)$
modified gravity%
\footnote{The Palatini gravity model discussed in this work does not introduce
extra degrees of freedom. Therefore, Ostrogradski-type~\cite{Woodard:2006nt}
instabilities do not occur despite the squared Ricci tensor~$Q$
in the action.%
} in the Palatini formalism, where $R$ is the Ricci scalar and $Q$
the squared Ricci tensor. In terms of the space-time metric~$g_{ab}$
and its determinant~$g=\text{det}\, g_{ab}$ the complete action
is given by
\begin{equation}
\mathcal{S}=\int d^{4}x\,\left[\sqrt{|g|}\frac{1}{2}f(R,Q)\right]+\mathcal{S}_{\text{mat}}[g_{ab,}\phi]\label{eq:P-fRQ-action}
\end{equation}
with the matter fields~$\phi$ in $\mathcal{S}_{\text{mat}}$ coupled
minimally to~$g_{ab}$. Furthermore, here the connection~$\Gamma_{bc}^{a}$
and the Ricci tensor~$R_{ab}$ are restricted to be symmetric,%
\footnote{See Ref.~\cite{Vitagliano:2010pq} for generalisations.%
} 
\begin{eqnarray}
R_{ab} & = & \Gamma_{ab,e}^{e}-\Gamma_{eb,a}^{e}+\Gamma_{ab}^{e}\Gamma_{fe}^{f}-\Gamma_{af}^{e}\Gamma_{eb}^{f}\label{eq:P-RicciTensor}\\
R & = & g^{ab}R_{ab},\,\,\,\,\,\,\,\,\,\,\, Q=R^{ab}R_{ab}=g^{ac}g^{bd}R_{ab}R_{cd}.\label{eq:P-Ricci-RQ}
\end{eqnarray}
In the metric formalism the connection would be identified with the
Levi-Civita connection
\begin{equation}
\Gamma_{bc}^{a}=\frac{1}{2}g^{ad}(g_{dc,b}+g_{bd,c}-g_{bc,d}),\label{eq:LC-metric}
\end{equation}
and the Ricci tensor and all derived quantities would be functionals
of the metric only. Differently, the Palatini formalism treats the
connection as an independent object in the beginning. Therefore, the
variational principle~$\delta\mathcal{S}=0$ yields two equations
of motion for gravity, one for the metric, 
\begin{equation}
f_{R}R_{m}^{\,\,\, n}+2f_{Q}R_{m}^{\,\,\, a}R_{a}^{\,\, n}-\frac{1}{2}\delta_{m}^{\,\,\, n}f=T_{m}^{\,\,\, n},\label{eq:P-EOM-1}
\end{equation}
 and another one for the connection,
\begin{equation}
\nabla_{a}\left[\sqrt{|g|}(f_{R}g^{mn}+2f_{Q}R^{mn})\right]=0,\label{eq:P-EOM-2}
\end{equation}
where~$\nabla_{a}$ denotes the covariant derivative in terms of
the yet unknown connection~$\Gamma_{bc}^{a}$ and~$f_{R,Q}$ are
the partial derivatives of~$f$ with respect to~$R$ and~$Q$,
respectively.

In the following we briefly summarise how to solve these equations,
and we point the reader to Refs.~\cite{Bauer:2010bu,Olmo:2009xy}
for a detailed exposition. Since we focus on the background cosmology,
the matter sector can be described by a perfect fluid with the energy-stress
tensor given by
\begin{equation}
T_{m}^{\,\,\, n}=(\rho+p)u_{m}u^{n}+(p-\Lambda)\delta_{m}^{\,\, n},\label{eq:StressTensor}
\end{equation}
where $\rho$ is the energy density and $p$ the pressure of ordinary
matter with the $4$-velocity vector $u_{m}$ normalised by $u_{m}u^{m}=-1$.
$\Lambda$ denotes the large CC vacuum energy density.

From Eq.~(\ref{eq:P-EOM-1}) two algebraic equations for the two
unknowns~$R$ and~$Q$ can be derived, the first one is the trace
of~(\ref{eq:P-EOM-1}) 
\begin{equation}
f_{R}R+2Qf_{Q}-2f=T,\label{eq:P-EOM-1-trace}
\end{equation}
where $T=T_{m}^{\,\, m}$. And the second equation follows from treating~(\ref{eq:P-EOM-1})
as a matrix equation for the Ricci tensor~$\hat{P}:=R_{m}^{\,\,\, n}$
in matrix form,%
\footnote{Here, $\hat{I}=\delta_{m}^{\,\, n}$ is the identity matrix, and $\hat{P}^{2}=R_{m}^{\,\, a}R_{a}^{\,\, n}$.%
}
\begin{equation}
(2f_{Q})^{2}\left(\hat{P}+\frac{1}{4}\frac{f_{R}}{f_{Q}}\,\hat{I}\right)^{2}=X^{2}\,\hat{I}+2f_{Q}(\rho+p)\,\widehat{u_{m}u^{n}}\label{eq:P-EOM-1-matrix}
\end{equation}
 with
\begin{equation}
X^{2}:=2f_{Q}(p-\Lambda)+f_{Q}\, f+\frac{1}{4}f_{R}^{2},\,\,\,\,\,\,\, Y:=X-c_{b}\cdot\sqrt{X^{2}-2f_{Q}(\rho+p)},\label{eq:X2Y-Def}
\end{equation}
The quadratic equation~(\ref{eq:P-EOM-1-matrix}) has the solutions
\begin{equation}
c_{a}\cdot2f_{Q}\left(\hat{P}+\frac{1}{4}\frac{f_{R}}{f_{Q}}\,\hat{I}\right)=X\,\hat{I}+Y\,\widehat{u_{m}u^{n}},\label{eq:P-EOM-1-sol}
\end{equation}
where $c_{a,b}=\pm1$ and the signs of the roots in $X$ and $Y$
follow from requiring that the final solution of the model solves
Eq.~(\ref{eq:P-EOM-1}). The trace of~(\ref{eq:P-EOM-1-sol}) yields
\begin{equation}
c_{a}(2f_{Q}R+2f_{R})=4X-Y,\label{eq:P-EOM-1-sol-trace}
\end{equation}
and at this point we simply eliminate all roots and $\pm1$ factors
by squaring Eq.~(\ref{eq:P-EOM-1-sol-trace}) twice, 
\begin{equation}
\left(\left(2f_{Q}R+2f_{R}\right)^{2}+8X^{2}+2f_{Q}(\rho+p)\right)^{2}=36X^{2}\left(2f_{Q}R+2f_{R}\right)^{2}.\label{eq:P-BigEq}
\end{equation}
Using~(\ref{eq:P-EOM-1-trace}) and~(\ref{eq:P-BigEq}), the Ricci
scalar~$R$ and the squared Ricci tensor~$Q$ can be determined
as algebraic functions of $\rho$, $p$ and~$\Lambda$. 

Next, we derive the connection~$\Gamma$ with the help of the auxiliary
metric~$h_{mn}$ defined by 
\begin{equation}
\sqrt{|h|}h^{mn}=\sqrt{|g|}g^{ma}\left(f_{R}\hat{I}+2f_{Q}\hat{P}\right)_{a}^{\,\, n},\label{eq:P-SigmaDef}
\end{equation}
which allows writing Eq.~(\ref{eq:P-EOM-2}) in the form
\begin{equation}
\nabla_{a}[\Gamma]\left[\sqrt{|h|}h^{mn}\right]=0.\label{eq:P-EOM-2-h}
\end{equation}
Since $\hat{P}$ follows from Eq.~(\ref{eq:P-EOM-1-sol}), one finds
after some algebra that $h_{mn}$ is related to the ``physical''
metric~$g_{mn}$ by the disformal transformation~\cite{Olmo:2009xy}
\begin{eqnarray}
h_{mn} & = & \Omega\left(g_{mn}-\frac{L_{2}}{L_{1}-L_{2}}u_{m}u_{n}\right),\,\,\,\,\,\,\, h^{mn}=\Omega^{-1}\left(g^{mn}+\frac{L_{2}}{L_{1}}u^{m}u^{n}\right),\label{eq:h_mn}
\end{eqnarray}
where 
\begin{equation}
\Omega:=\sqrt{|L_{1}(L_{1}-L_{2})|},\,\,\,\,\,\,\,\, L_{1}:=c_{a}X+\frac{1}{2}f_{R},\,\,\,\,\,\,\,\, L_{2}:=c_{a}Y.\label{eq:OmegaL1L2Def}
\end{equation}
Finally, the Palatini connection~$\Gamma$ is just the solution of~(\ref{eq:P-EOM-2-h}),
i.e.\ the Levi-Civita connection of $h_{mn}$,
\begin{equation}
\Gamma_{bc}^{a}[h]=\frac{1}{2}h^{ad}(h_{dc,b}+h_{bd,c}-h_{bc,d})\label{eq:LC-Con-h}
\end{equation}
with~$h_{mn}$ given in~(\ref{eq:h_mn}). This result solves Eq.~(\ref{eq:P-EOM-2})
and eventually defines the Ricci tensor and all quantities derived
from it. Note that $h_{mn}$ does not contain space-time derivatives
because it is a function of only $g_{ab}$, $\rho$, $p$ and $\Lambda$.
Hence,~$R_{ab}$, $R$ and the other quantities calculated according
to Eqs.~(\ref{eq:P-RicciTensor}) and~(\ref{eq:P-Ricci-RQ}) contain
derivatives of at most second order of the metric~$g_{ab}$. Consequently,
in this modified gravity theory in the Palatini formalism one has
to solve the second order differential equation
\begin{equation}
R[\Gamma]=R(\rho,p,\Lambda),\label{eq:Diff-Equ-R}
\end{equation}
where the right-hand side follows from the algebraic solution of Eqs.~(\ref{eq:P-EOM-1-trace})
and~(\ref{eq:P-BigEq}), whereas the left-hand side is given in Eq.~(\ref{eq:P-Ricci-RQ})
using~$\Gamma$ from Eq.~(\ref{eq:LC-Con-h}). In the next section,
we will apply this procedure to the CC filter model.

\section{CC filter model}

\label{sec:CC-Filter}In Ref.~\cite{Bauer:2010bu} a modified gravity
model was constructed that filters out vacuum energy contributions
(i.e.\ everything with the EOS $p=-\rho$) to the CC. It is defined
by the following action functional
\begin{equation}
f(R,Q)=\kappa R+z,\,\,\,\, z:=\beta\,\left(\frac{R^{\frac{2}{3}}}{B}\right)^{m},\,\,\,\, B:=R^{2}-Q,\label{eq:fRQ-ansatz}
\end{equation}
with the dimensionful parameters~$\kappa$ and $\beta$ and the positive
number~$m$. Note that in this setup, the term~$z$ is never considered
to be a little correction to $\kappa R$, instead both terms in~$f$
act in collaboration. From here one, we will use~$B$ instead of~$Q$.
With the partial derivatives of~$f$, $f_{R}=(\kappa R+\frac{2}{3}mz)/R-2f_{Q}R$,
$f_{Q}=mz/B$, and the trace~$T=-4\Lambda+3p-\rho$ of the energy-stress
tensor~(\ref{eq:StressTensor}), Eq.~(\ref{eq:P-EOM-1-trace}) becomes
\begin{equation}
\gamma z=\kappa R-4\Lambda+3p-\rho,\,\,\,\,\,\gamma:=-2-\frac{4}{3}m,\label{eq:z}
\end{equation}
while Eq.~(\ref{eq:P-BigEq}) can be written as 
\begin{eqnarray}
0 & = & \kappa R+r-\frac{2}{9}mz\left(\frac{B}{R^{2}}\right)\left[1+\frac{3}{mz}\left(3\kappa R+2r\right)+\frac{9}{4(mz)^{2}}\left((\kappa R)^{2}-r^{2}\right)\right]\label{eq:SeqZeroRelax}\\
 & + & \frac{8}{27}mz\left(\frac{B}{R^{2}}\right)^{2}\left(1+\frac{3\kappa R}{2(mz)}\right)^{2}.\nonumber 
\end{eqnarray}
In the last equation we introduced the variable $r:=\rho+p$, which
vanishes identically for CC contributions.

For matter energy densities~$\rho$ and curvatures $\kappa R$ smaller
in magnitude than the large CC, the term $\gamma z\approx-4\Lambda$
in~(\ref{eq:z}) is approximately constant and Eq.~(\ref{eq:SeqZeroRelax})
determines the curvature as a function of $r$ alone. Therefore, contributions
to the CC and shifts in the vacuum energy density are filtered out.
This limit of constant $z=\mathcal{O}(\Lambda)$ covers most of the
radiation epoch, the matter era and the late-time de~Sitter phase,
and it has been analysed analytically in Ref.~\cite{Bauer:2010bu}
by neglecting strongly suppressed terms like $\mathcal{O}(r/z)$.

In the following we complement our previous results by investigating
the very early universe, where~$r$ may be of the same order as~$\Lambda$
or above. Consequently, $z$ in Eq.~(\ref{eq:z}) is not constant
anymore and all terms in Eq.~(\ref{eq:SeqZeroRelax}) must be taken
into account. It is convenient to work with dimensionless variables,
where energy densities and the curvature term~$\kappa R$ are normalised
by the cosmological constant~$(4\Lambda)$,
\begin{equation}
y:=\frac{\kappa R}{4\Lambda},\,\,\,\, x:=\frac{r}{4\Lambda}=\frac{\rho+p}{4\Lambda}.\label{eq:xy-Def}
\end{equation}
Moreover, let us define
\begin{equation}
m_{z}:=\frac{m\cdot z}{4\Lambda}=\frac{m}{-\gamma}\left(1-y-x\cdot\omega_{m}\right)\,\,\,\,\text{with}\,\,\,\,\omega_{m}:=\frac{3\omega-1}{\omega+1},\label{eq:mz-Def}
\end{equation}
where $\omega=p/\rho$ denotes the matter EOS. Finally, we introduce
the variable
\begin{equation}
W:=\frac{R^{2}}{B}=y^{\frac{4}{3}}\left(1-y-x\cdot\omega_{m}\right)^{\frac{1}{m}}\cdot\delta,\label{eq:W-Def}
\end{equation}
where~$B$ was eliminated on the right-hand side by using the definition
of~$z$ in Eq.~(\ref{eq:fRQ-ansatz}). The constant~$\delta$ is
completely fixed by the model parameters and the large vacuum energy
density~$\Lambda$,
\begin{equation}
\delta:=\left(\frac{2m}{-9\gamma}\right)\left(\frac{4\Lambda}{\rho_{e}}\right)^{\frac{7}{3}},\,\,\,\,\text{where}\,\,\,\,\left(\rho_{e}\right)^{\frac{7}{3}}:=\kappa^{\frac{4}{3}}\left(\frac{2}{9}m\frac{4\Lambda}{-\gamma}\right)\left(\frac{-\gamma\beta}{4\Lambda}\right)^{\frac{1}{m}}.\label{eq:delta-Def}
\end{equation}
According to our earlier results in Ref.~\cite{Bauer:2010bu}, $\delta$
is very large in magnitude because~$\rho_{e}$ is of the order of
the late-time critical energy density in the limit $\gamma z\approx-4\Lambda$.
Expressed in terms of the dimensionless variables the main equation~(\ref{eq:SeqZeroRelax})
reads
\begin{eqnarray}
0 & = & y+x-\frac{2}{9}\frac{m_{z}}{W}\left[1+\frac{3(3y+2x)}{m_{z}}+\frac{9(y^{2}-x^{2})}{4m_{z}^{2}}\right]+\frac{8}{27}\frac{m_{z}}{W^{2}}\left[1+\frac{3y}{2m_{z}}\right]^{2}.\label{eq:MainEq-xy}
\end{eqnarray}
For the late-time solutions mentioned above we find $m_{z}=-m/\gamma=\mathcal{O}(1)$,
$|y|,|x|\ll1$ and $|W|\gg1$. Therefore, Eq.~(\ref{eq:MainEq-xy})
simplifies considerably,
\begin{equation}
y+x-\frac{2}{9}\frac{m_{z}}{W}=0,\label{eq:MainEq-xy-late}
\end{equation}
indicating $\kappa R\rightarrow\rho_{e}$ for vanishing matter $r\rightarrow0$,
which justifies introducing~$\rho_{e}$ in Eq.~(\ref{eq:delta-Def}).

In the next section, we solve the complete equation~(\ref{eq:MainEq-xy})
requiring a numerical treatment. In practise, we work along the following
procedure. First, we consider a spatially flat cosmological background
with the expansion described by the scale factor~$a(t)$ as a function
of cosmological time. In Cartesian coordinates ($i=1,2,3$) the non-zero
components of the metric are given by $g_{00}=-1$, $g_{ii}=a^{2}(t)$,
and in the stress tensor we use~$u_{m}=\delta_{m}^{0}$ . Hence,
the auxiliary metric in~(\ref{eq:h_mn}) has only the diagonal components
\begin{equation}
h_{00}=\Omega\left(\frac{-L_{1}}{L_{12}}\right),\,\,\, h_{ii}=\Omega\, a^{2}(t),\,\,\, L_{12}:=L_{1}-L_{2},
\end{equation}
which determine via (\ref{eq:LC-Con-h}) the symmetric components
of the Palatini connection
\begin{eqnarray}
\Gamma_{00}^{0} & = & \frac{\dot{x}}{4}\left(3\frac{L_{1}^{\prime}(x)}{L_{1}}-\frac{L_{12}^{\prime}(x)}{L_{12}}\right)\nonumber \\
\Gamma_{ii}^{0} & = & \frac{a^{2}}{L_{1}}\left(\frac{\dot{a}}{a}L_{12}+\frac{\dot{x}}{4}\left[L_{12}^{\prime}(x)+\frac{L_{12}L_{1}^{\prime}(x)}{L_{1}}\right]\right)\nonumber \\
\Gamma_{i0}^{i} & = & \frac{\dot{a}}{a}+\frac{\dot{x}}{4}\left(\frac{L_{12}^{\prime}(x)}{L_{12}}+\frac{L_{1}^{\prime}(x)}{L_{1}}\right).\label{eq:Palatini-Gammas}
\end{eqnarray}
Here, the overdot indicates a derivative with respect to the time
coordinate~$t$, whereas the prime denotes the partial derivative
with respect to~$x$. For simplicity we consider matter with a constant
EOS~$\omega$, which obeys the standard conservation equation~$\dot{\rho}+3\frac{\dot{a}}{a}\rho(1+\omega)=0$
implying~$\rho\propto a^{-3(1+\omega)}$ and $\dot{x}=-3\frac{\dot{a}}{a}(1+\omega)x$
via $x=(1+\omega)\rho/(4\Lambda)$. Consequently, the Ricci scalar~$R(t)$
calculated via~(\ref{eq:P-Ricci-RQ}) from the connection~(\ref{eq:Palatini-Gammas})
is a function of $a$, $\dot{a}$ and~$\ddot{a}$, whereas the algebraic
solution $y(x)=\kappa R/(4\Lambda)$ of Eq.~(\ref{eq:MainEq-xy})
is only a function of~$a$. For obtaining~$a(t)$ we solve numerically
the differential equation~(\ref{eq:Diff-Equ-R}),
\begin{equation}
\frac{R(t)}{H_{0}^{2}}=K\cdot y(x(t))\,\,\,\,\text{with}\,\,\,\, K:=\frac{4\Lambda}{\kappa H_{0}^{2}},\label{eq:ReqKy}
\end{equation}
where~$H_{0}=\frac{\dot{a}}{a}(t_{0})$ is the Hubble rate at the
arbitrary initial time~$t_{0}$. In Ref.~\cite{Bauer:2010bu} we
found~$(-\kappa)$ to be of the order of the inverse Newton's constant~$G$
indicating~$\kappa H_{0}^{2}\sim\rho_{e}$ by identifying~$H_{0}$
with the present Hubble rate and using the late-time solution~(\ref{eq:MainEq-xy-late}).
In this case, $K$ is roughly the square root of~$\delta$ in~(\ref{eq:delta-Def}),
its exact value is fixed when we specify~$H_{0}$ and $x(t_{0})$
at the initial%
\footnote{Alternatively, one could start with given values of~$\kappa$ and
$x(t_{0})$ and infer~$H_{0}$ from it. This is analogous to finding~$H_{0}$
from the Friedmann equation $3H_{0}^{2}=8\pi G\cdot\rho(t_{0})$ in
general relativity.%
} time~$t_{0}$. Numerically, we first solve the equation%
\footnote{We use this equation in terms of arbitrary~$t$ to determine the
correct choice of~$c_{a,b}$ in~(\ref{eq:X2Y-Def}) and~(\ref{eq:OmegaL1L2Def})
after Eq.~(\ref{eq:ReqKy}) has been solved for~$a(t)$.%
} $R^{2}/B=W(y(x))$ for~$\ddot{a}(t_{0})$, where $R^{2}/B$ has
to be calculated from~(\ref{eq:Palatini-Gammas}) at $t=t_{0}$ and
$x=x(a_{0})$. Then, with~$\ddot{a}(t_{0})$ the values of $R(t_{0})$
and $y(x(t_{0}))$ in Eq.~(\ref{eq:ReqKy}) are known, which yields
the numerical value of~$K$ and Eq.~(\ref{eq:ReqKy}) can be solved
to determine the evolution of the scale factor~$a(t)$ for all times~$t$.

\section{Solutions}

\label{sec:Solutions}Here, we study the cosmological evolution resulting
from the differential equation~(\ref{eq:ReqKy}). Thereby we use
the solutions to the main equation~(\ref{eq:MainEq-xy}), which relate
the curvature~$y$ to the matter energy density and pressure in~$x$
as given in~(\ref{eq:xy-Def}). We do not restrict the magnitudes
of~$x$ and~$y$, however, in order to be roughly compatible with
our existent universe, we consider only solutions $y(x)$ with both
a final accelerating phase, $x\rightarrow0$, and a high energy decelerating
epoch, $y\approx-x$. Moreover, for the numerical examples below we
set $m=1/3$ in Eq.~(\ref{eq:fRQ-ansatz}), because this parameter
has only little influence on the qualitative behaviour of the solutions.
With these preliminaries, there are four qualitatively different cases
to study corresponding to $\Lambda\lessgtr0$ and $\delta\lessgtr0$.
Note that the sign of~$\delta$ can be adjusted by a suitable choice
of the parameter~$\beta$ in the action functional~(\ref{eq:fRQ-ansatz}).
\begin{enumerate}
\item The absence of an initial singularity can be seen best in the case~$\delta>0$
and~$\Lambda<0$ implying~$\rho_{e}<0$ via Eq.~(\ref{eq:delta-Def})
and~$\gamma<0$ from~(\ref{eq:z}). In this case the matter variable~$x=r/(4\Lambda)$
is negative because of~$r=\rho+p\ge0$. Numerically, we do not find
solutions for~$y<0$, that satisfy our requirements given above.
This can be inferred also from Eq.~(\ref{eq:MainEq-xy-late}), where
all terms would be negative. On the other hand, for~$y>0$ (or~$\kappa R<0$)
we find Eq.~(\ref{eq:MainEq-xy-late}) as a good low-energy description
with $y(x\rightarrow0)\rightarrow\rho_{e}/(4\Lambda)$ and $y\approx-x$
for~$|\rho_{e}|\ll|x|\ll1$. However, before~$y\rightarrow1$, the
matter variable~$|x|$ becomes smaller again and eventually vanishes
for finite~$y<1$ as illustrated in Fig.~\ref{fig:Sol-1a-YofX}.
\begin{figure}
\begin{centering}
\includegraphics[width=1\textwidth]{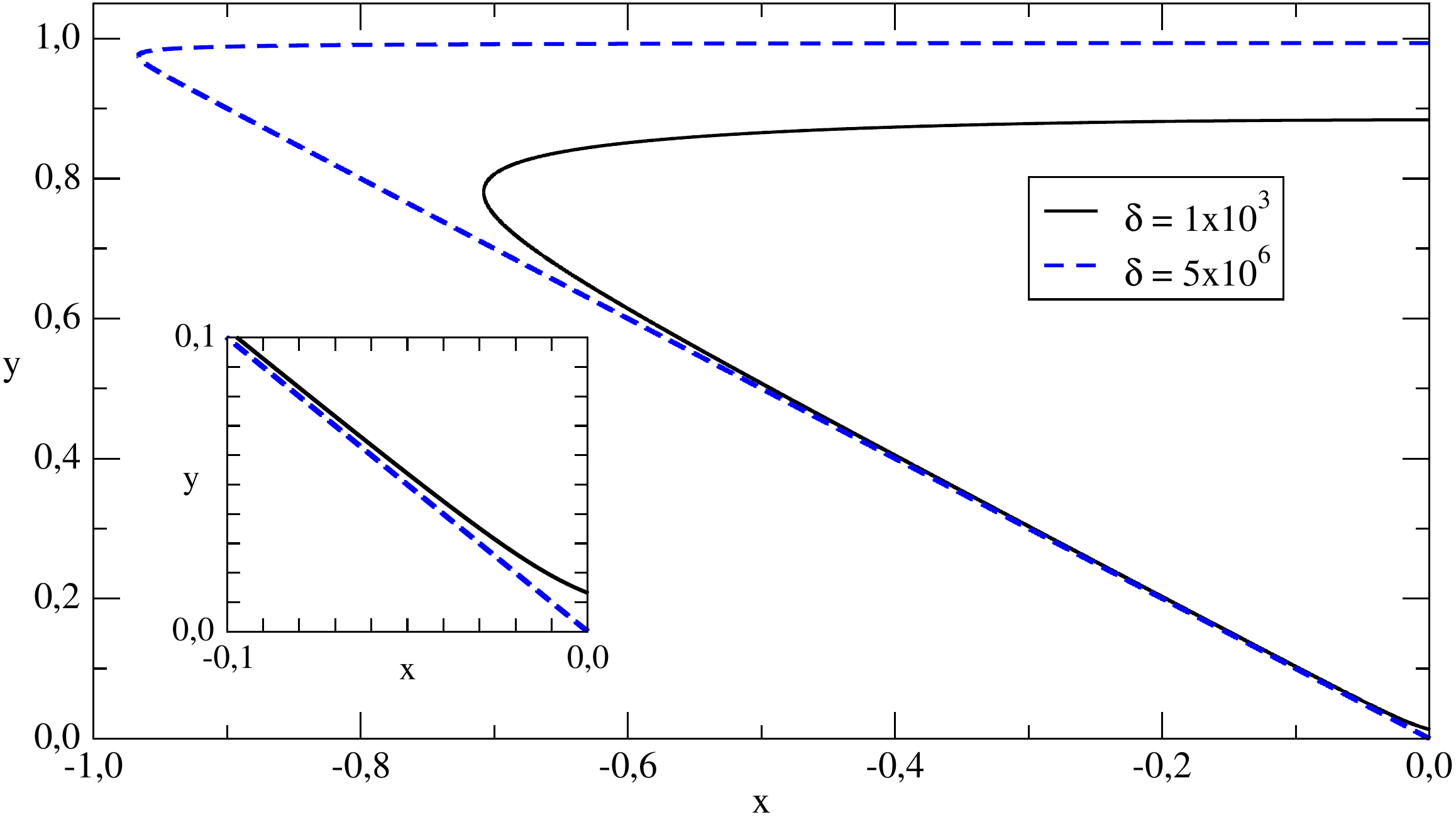}
\par\end{centering}

\caption{\label{fig:Sol-1a-YofX}The two branches of the function~$y(x)$
from Eq.~(\ref{eq:MainEq-xy}) relating the curvature term~$y=\kappa R/(4\Lambda)$
to the matter variable~$x=r/(4\Lambda)$ for the case~$\delta>0$,
$\Lambda<0$, EOS~$\omega=\frac{1}{3}$.}
\end{figure}
 As a result, $|x|\propto a^{-3(\omega+1)}$ and $y$ are bounded
from above signalling the absence of the initial singularity, where
$x$ or~$y$ would diverge. Analytically, this boundedness can be
understood from Eqs.~(\ref{eq:mz-Def}) and~(\ref{eq:W-Def}) indicating
that for sufficiently large values of~$y$ or~$|x|$ the variables~$W$
and~$m_{z}$ vanish. Since they appear in the denominators in our
main equation~(\ref{eq:MainEq-xy}), it is clear that neither $|x|$
nor~$y$ can become larger than~$1$. After solving the differential
equation~(\ref{eq:ReqKy}) for the scale factor~$a(t)$, we observe
that not the whole phase-space curve~$y(x)$ in Fig.~\ref{fig:Sol-1a-YofX}
is realised, but only a smaller part of it. This behaviour can be
explained by observing that the function~$L_{1}-L_{2}$ vanishes
at $x\approx-0,043$ as shown in Fig.~\ref{fig:Sol-1a-L1L1mL2}.
\begin{figure}
\begin{centering}
\includegraphics[width=1\textwidth]{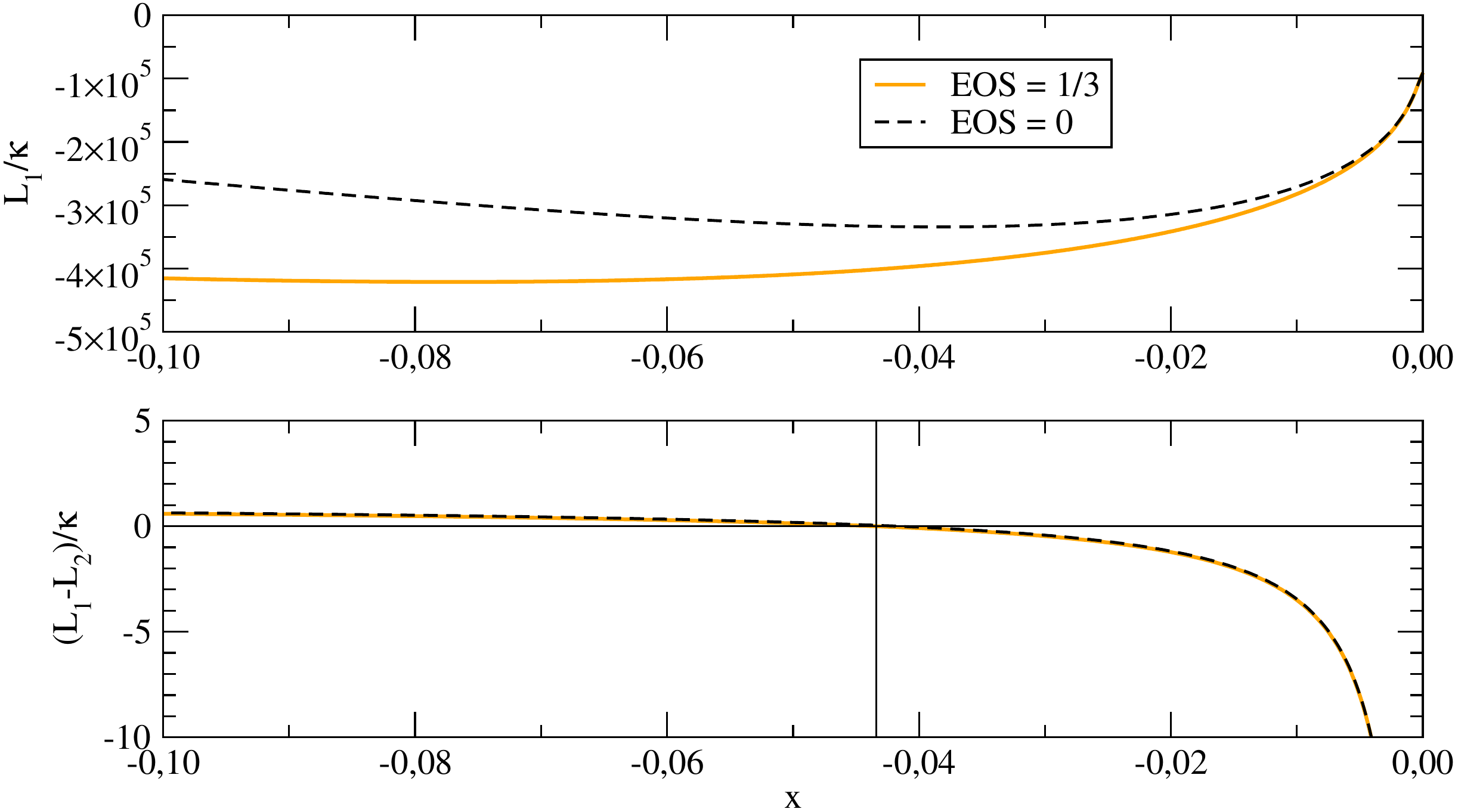}
\par\end{centering}

\caption{\label{fig:Sol-1a-L1L1mL2}The functions~$L_{1}(x)$ and~$L_{12}(x)$
in the connection components in Eqs.~(\ref{eq:Palatini-Gammas})
for the case~$\delta=5\cdot10^{6}$, $\Lambda<0$ and two values
of the EOS~$\omega=\frac{1}{3},\,0$. Notice the zero of~$L_{12}$
at $x\approx-0,043$, where the cosmic bounce occurs.}
\end{figure}
 Since~$L_{1}-L_{2}$ appears in the denominator of the connection
components~(\ref{eq:Palatini-Gammas}) and therefore in the Ricci
scalar~$R(t)$, the equation~(\ref{eq:ReqKy}) can be fulfilled
only if~$\dot{a}/a\propto\dot{x}/x$ vanishes at the zero of~$L_{1}-L_{2}$
because~$y(x(t))$ on the right-hand side of~(\ref{eq:ReqKy}) is
always finite. Accordingly, $\dot{a}=0$ indicates the occurrence
of a cosmic bounce, which is shown in the plots for the scale factor~$a(t)$,
the Hubble rate~$H=\dot{a}/a$ and the deceleration~$q(t)$ in Fig.~\ref{fig:Sol-1a-aHq}.
\begin{figure}
\begin{centering}
\includegraphics[width=1\textwidth]{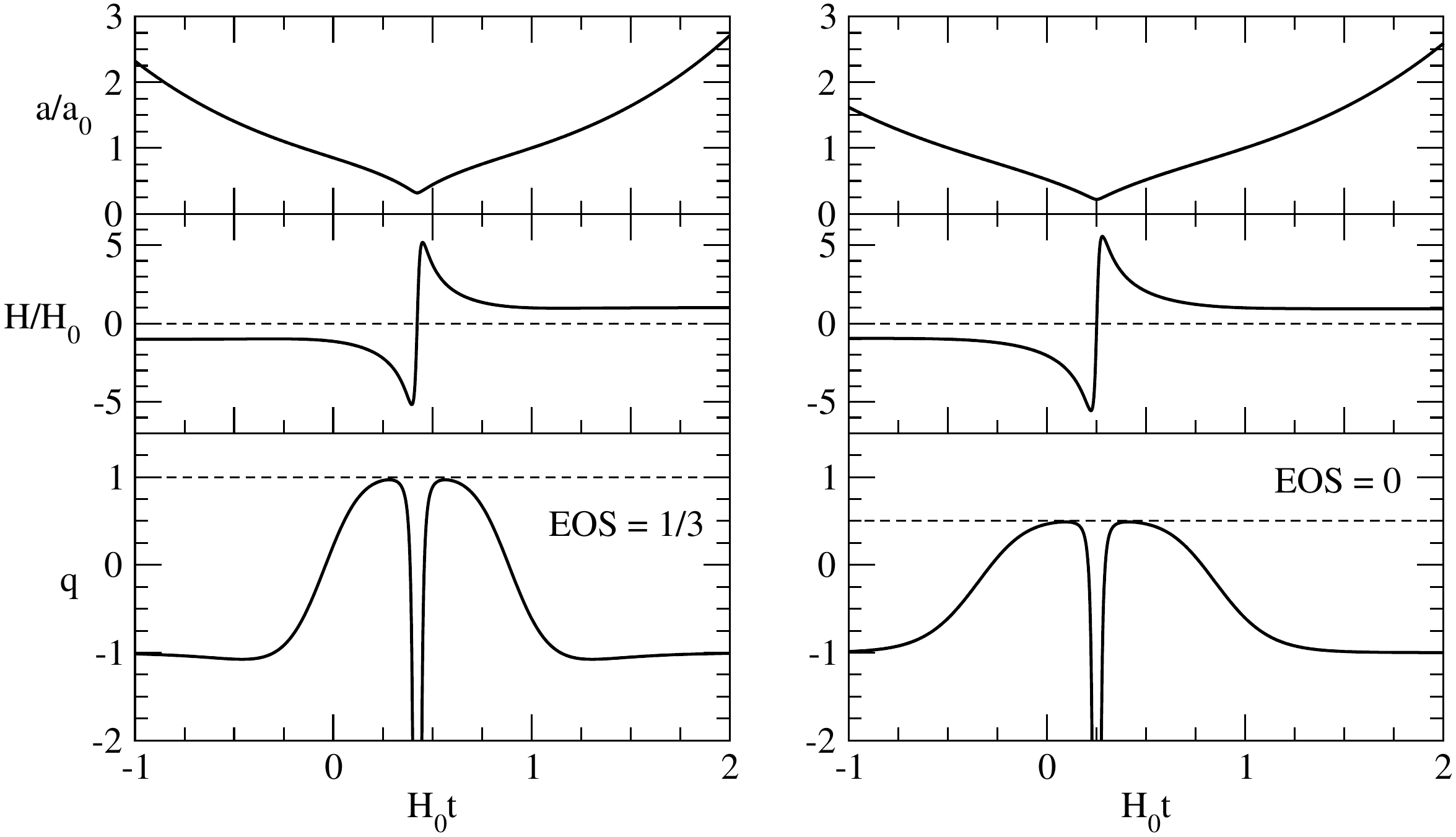}
\par\end{centering}

\caption{\label{fig:Sol-1a-aHq}The numerical solutions from Eq.~(\ref{eq:ReqKy})
for the scale factor~$a(t)$, the Hubble rate~$H(t)$ and the deceleration
factor~$q=-\frac{\ddot{a}}{a}/H^{2}$ for the case~$\delta=5\cdot10^{6}$,
$\Lambda<0$ described in Fig.~\ref{fig:Sol-1a-L1L1mL2}.}
\end{figure}
 Also, we can confirm the analytical results from Ref.~\cite{Bauer:2010bu},
where it was found that at early times the universe behaves like a
radiation- or dust-dominated cosmos depending on the matter EOS~$\omega$.
Thus for $\omega=\frac{1}{3}$ or $0$ the deceleration~$q$ approaches
the value $1$ or~$\frac{1}{2}$, respectively. In this era, $y\approx-x$
is a good approximation. Finally, at late times ($x\rightarrow0$)
we find a de~Sitter cosmos with $y\rightarrow\rho_{e}/(4\Lambda)$.
\item The second case we consider is~$\delta<0$ and~$\Lambda>0$ implying~$x>0$
and~$\rho_{e}<0$. Here, the only reasonable solution requires~$y,\kappa R<0$.
In contrast to the first case, the variables~$x$ and~$|y|$ are
not bounded from above because~$W$ and~$m_{z}$ in Eqs.~(\ref{eq:mz-Def})
and~(\ref{eq:W-Def}) do not vanish. Instead we find~$y\rightarrow-x$
for large~$x\rightarrow\infty$, see Fig.~\ref{fig:Sol-4a-YofX-L1L2mL2}.
\begin{figure}
\begin{centering}
\includegraphics[width=1\textwidth]{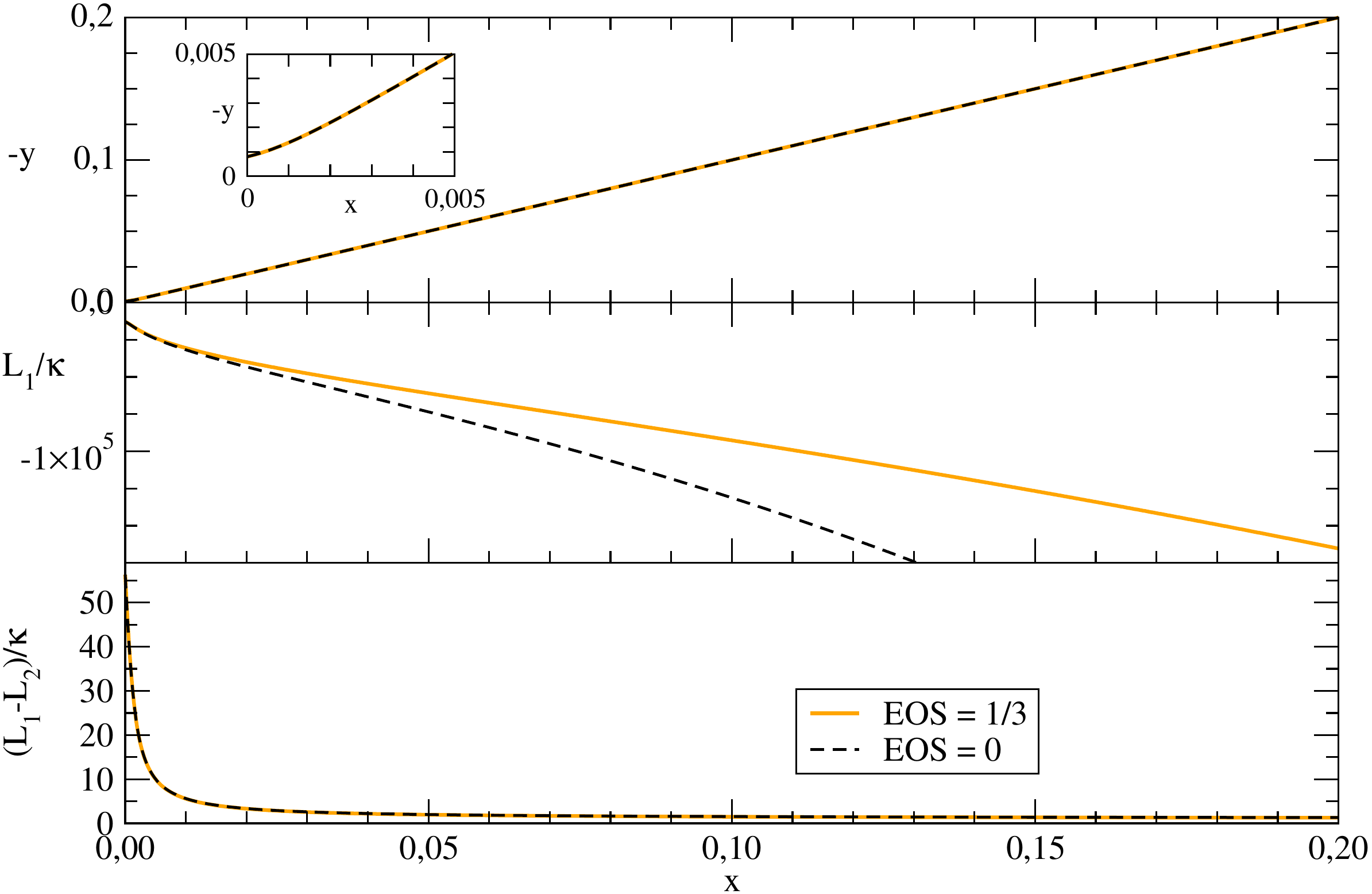}
\par\end{centering}

\caption{\label{fig:Sol-4a-YofX-L1L2mL2}The solution~$y(x)$ from Eq.~(\ref{eq:MainEq-xy})
and the functions~$L_{1}(x)$ and~$L_{12}(x)$ in the connection
in Eqs.~(\ref{eq:Palatini-Gammas}) for the case~$\delta=-5\cdot10^{6}$,
$\Lambda>0$ and two values of the EOS~$\omega=\frac{1}{3},\,0$.}
\end{figure}
 Also, the functions~$L_{1}$ and~$L_{1}-L_{2}$ do not vanish in
this region. Nevertheless, the cosmic evolution inferred from the
numerical solution of~(\ref{eq:ReqKy}) shows a cosmic bounce, where
the scale factor~$a(t)$ and the matter and curvature variables~$x,|y|$
remain finite, see Fig.~\ref{fig:Sol-4a-aHq}.
\begin{figure}
\begin{centering}
\includegraphics[width=1\textwidth]{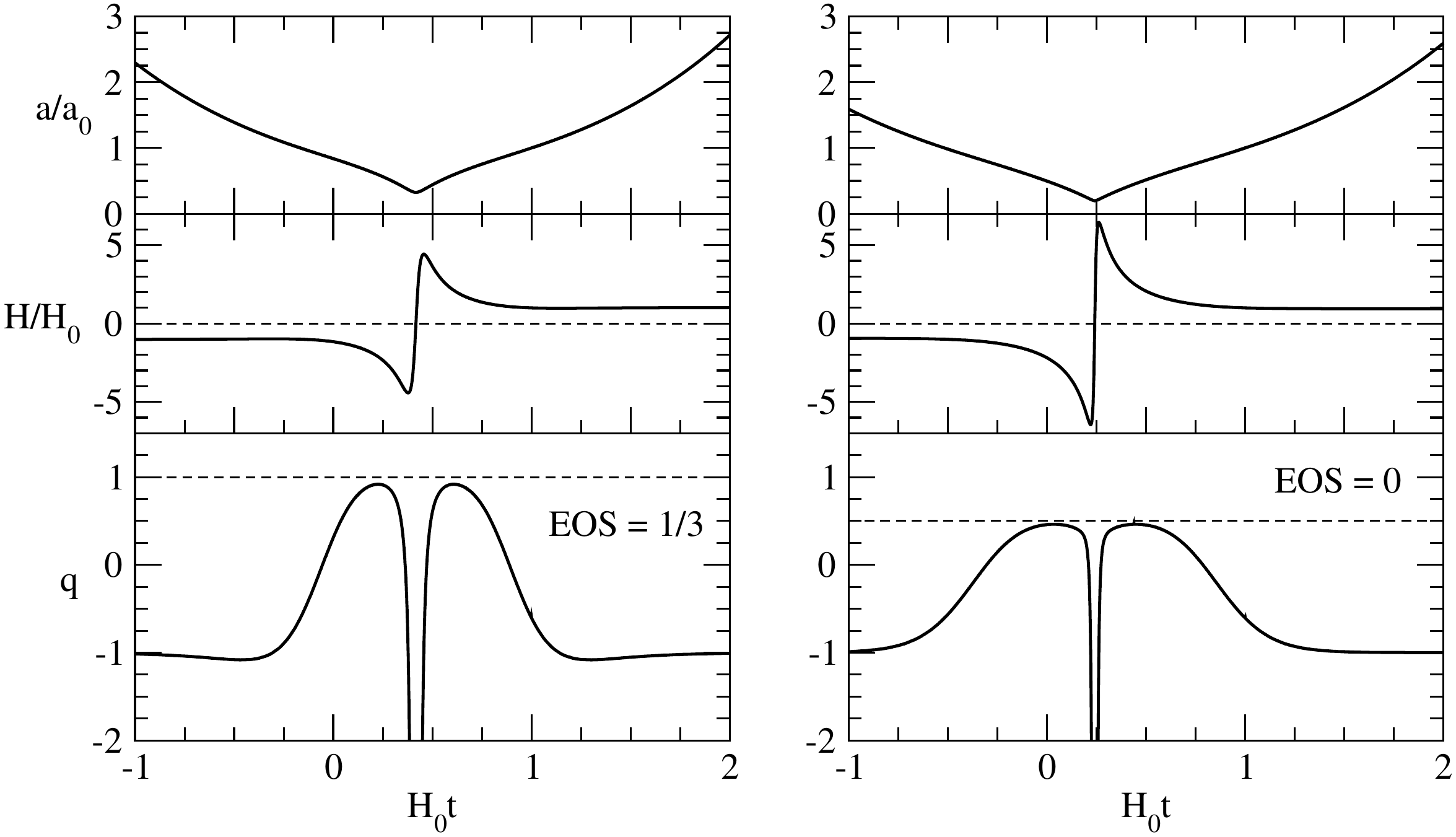}
\par\end{centering}

\caption{\label{fig:Sol-4a-aHq}The numerical solutions from Eq.~(\ref{eq:ReqKy})
for the scale factor~$a(t)$, the Hubble rate~$H(t)$ and the deceleration
factor~$q=-\frac{\ddot{a}}{a}/H^{2}$ for the case~$\delta=-5\cdot10^{6}$,
$\Lambda>0$ described in Fig.~\ref{fig:Sol-4a-YofX-L1L2mL2}.}
\end{figure}
 As before, the universe shows a radiation- or matter-dominated expansion
behaviour with~$y\approx-x$ and a final de~Sitter phase ($y\rightarrow\rho_{e}/(4\Lambda)$).
\item The third possibility is characterised by~$\delta<0$ and~$\Lambda<0$
indicating $x<0$ and $\rho_{e}>0$. We did not find a reasonable
solution in this case, although for~$y>0$ a $y\approx-x$ phase
exists, but it cannot end in a low-energy accelerating phase with~$x\rightarrow0$
because all terms in Eq.~(\ref{eq:MainEq-xy}) are positive in this
limit. Hence, we do not pursue this solution any further.
\item Finally, the last case with $\delta>0$ and $\Lambda>0$ implies $x>0$,
and for $y<0$ we find a radiation/matter phase described by $y\approx-x$
for~$x\rightarrow\infty$. Similar to the third case, $\rho_{e}>0$
does not seem to allow a late-time de~Sitter solution, however, the
last term~$\propto W^{-2}$ in Eq.~(\ref{eq:MainEq-xy}) is positive
giving rise to an accelerating solution with~$x\rightarrow0$. This
cosmos is, however, quite exotic, because there is a contraction phase
between the decelerating and the final de~Sitter phase. For completeness
the expansion behaviour of this interesting but unrealistic solution
is shown in Fig.~\ref{fig:Sol-2a-aHq-YofX}.
\begin{figure}
\begin{centering}
\includegraphics[width=1\textwidth]{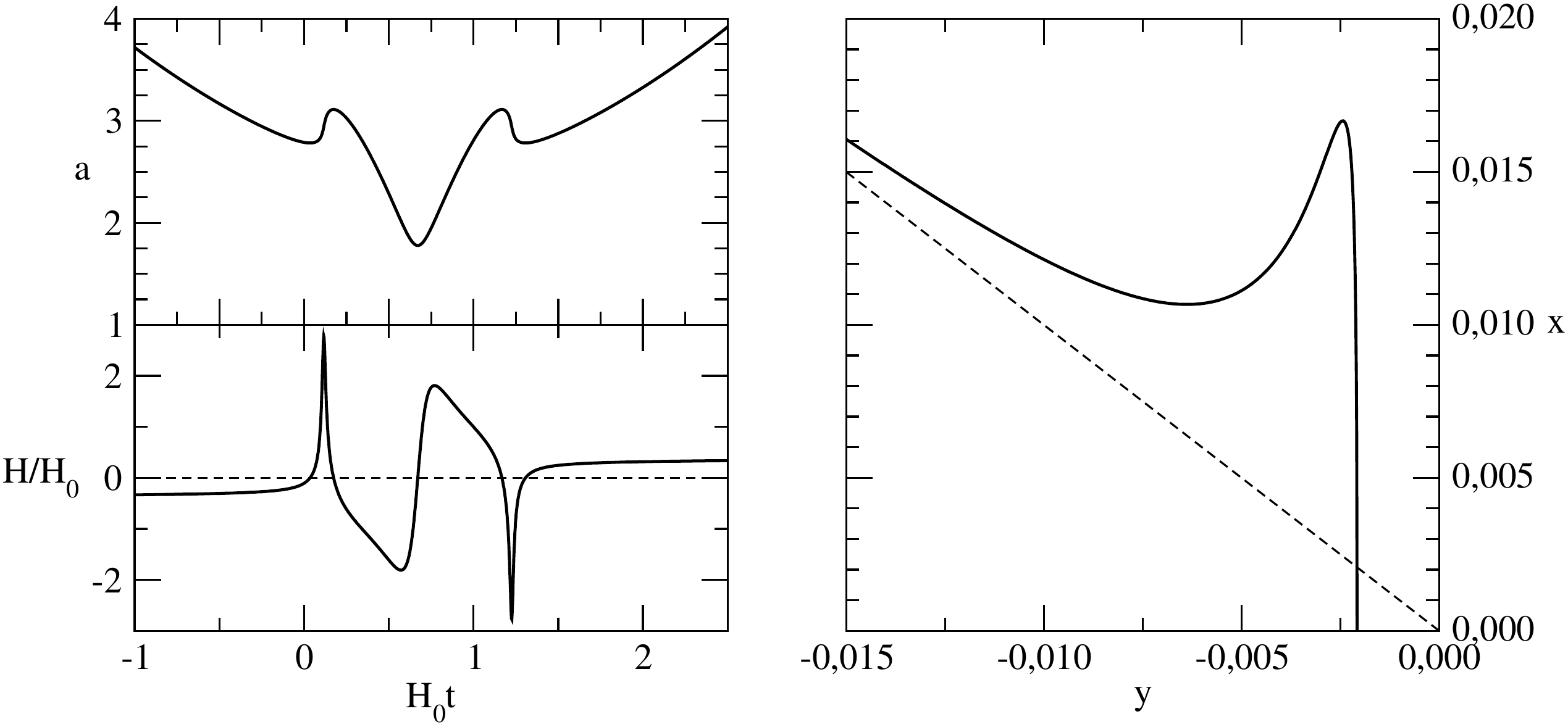}
\par\end{centering}

\caption{\label{fig:Sol-2a-aHq-YofX}The solution~$x(y)$ from Eq.~(\ref{eq:MainEq-xy}),
the scale factor~$a(t)$ and the Hubble rate~$H(t)$ from the numerical
solution of Eq.~(\ref{eq:ReqKy}) for the exotic case $\delta=5000$,
$\Lambda>0$. Notice the contraction phase between the decelerating
and accelerating eras.}
\end{figure}

\end{enumerate}
In the examples shown in this section we used relatively small values
of~$|\delta|$ for numerical reasons. According to its definition
in Eq.~(\ref{eq:delta-Def}), $\delta\sim(\Lambda/\rho_{e})^{7/3}$
is related to the ratio of the large CC and the late-time energy density
implying an enormous magnitude for~$\delta$. However, this does
not correspond to unreasonable values for the parameter~$\beta$
in the action~(\ref{eq:fRQ-ansatz}), instead a range of common values
can be obtained depending on~$m$ and~$\kappa$. Moreover, in this
work we have normalised energy densities like~$r$ and $\kappa R$
by the constant~$\Lambda$ such that our results are independent
of the numerical values of~$\kappa$ and~$\Lambda$. For details
on~$\beta$ and the relation between~$\kappa$ and Newton's constant
we refer the reader to Ref.~\cite{Bauer:2010bu}.

In summary, we found cosmological solutions for both signs of the
large CC~$\Lambda$, where the initial big bang singularity is replaced
by a cosmic bounce with finite curvature and finite matter energy
density. After the bounce these universes behave like being radiation
or matter dominated, where the deceleration~$q$ is close to~$1$
or~$\frac{1}{2}$, respectively. Eventually, when the matter component
is sufficiently diluted the final de~Sitter era begins, with an effective
vacuum energy density~$\sim\rho_{e}$ much smaller in magnitude than
the large CC~$\Lambda$. The latter property and the absence of an
initial cosmological singularity are the main benefits of the CC filter
model.

\section{Conclusions}

\label{sec:Conclusions}The filter model for a large CC in the context
of Palatini modified gravity~\cite{Bauer:2010bu} makes gravity insensitive
to vacuum energy density contributions. In other words, the curvature
is dominated by ordinary matter, whereas energy sources with the EOS
of a CC do not contribute at leading order despite having a much larger
energy density at late times.

In this work, we have explored the CC filter model in the context
of the very early universe, where in classical theories one usually
expects to find an initial big bang singularity at very high energies.
However, in our setup we observed that the curvature singularity is
replaced by a cosmic bounce with the matter energy density, the scale
factor and the curvature remaining finite. In the case shown in Fig.~\ref{fig:Sol-1a-YofX}
the boundedness can be seen already from the solution of the algebraic
equation~(\ref{eq:MainEq-xy}), whereas the other cases show the
same feature after solving the differential equation~(\ref{eq:ReqKy}).
Moreover, our numerical results also confirm the analytical study
of the late-time evolution in Ref.~\cite{Bauer:2010bu}.

The absence of an initial space-time singularity could mean that our
classical model for gravity does not require an ultraviolet completion
from a theory of quantum gravity. However, in the light of the exotic
form of the gravity action functional~(\ref{eq:fRQ-ansatz}) it is
unlikely that this setup is a fundamental theory. On the other hand,
we may follow the argumentation in Ref.~\cite{Olmo:2008nf}, where
Palatini gravity models are proposed as effective continuum descriptions
of loop quantum cosmology. In both approaches the regularisation of
the initial singularity does not require the introduction of new degrees
of freedom making them minimal solutions to this problem.

Independent of its theoretical origin, the CC filter model has the
capability to handle two fundamental problems of gravity, the old
CC problem and the avoidance of an initial singularity. Therefore,
it is an attractive scenario for further investigations. In addition
to the qualitative features discussed so far, it will be interesting
to study more aspects of the CC filter model in the future, e.g.\
black hole or star interiors or anisotropies and inhomogeneities in
cosmology.

\subsection*{Acknowledgements}

This work has been supported in part by MEC and FEDER under project
FPA2010-20807, by the Spanish Consolider-Ingenio 2010 program CPAN
CSD2007-00042 and by DIUE/CUR Generalitat de Catalunya under project
2009SGR502.

{\small \bibliographystyle{utphys}
\bibliography{Palatini-Bounce}
}
\end{document}